\documentstyle[aps,epsf,twocolumn]{revtex}
\begin{document}
\title{Odd numbers of photons and teleportation}
\author{S.J. van Enk\\
Bell Labs, Lucent Technologies
600-700 Mountain Ave\\
Murray Hill NJ 07974}
\maketitle
\abstract{Several teleportation protocols, namely those using entangled coherent states, entangled squeezed states, and the single-photon EPR state, are all shown to be particular instances of a more general scheme that relies on the detection of an odd number of photons.}
\medskip
\section{Introduction}
For any two pure states $|u\rangle$ and $|v\rangle$ whose inner product $\langle u|v\rangle$ is real, the states
\begin{eqnarray}\label{ent1}
|\Psi_-\rangle&=&|u\rangle|u\rangle-|v\rangle|v\rangle\nonumber\\
|\Phi_-\rangle&=&|u\rangle|v\rangle-|v\rangle|u\rangle
\end{eqnarray}
(leaving out the normalization factors for now) contain exactly one ebit\cite{ent} of entanglement, even if $|u\rangle$ and $|v\rangle$ are not orthogonal. This was noted in Ref.~\cite{hirota} for general states, and for $|u\rangle$ and $|v\rangle$ coherent states of the electromagnetic (EM) field with opposite phases in particular.
Such states, then, can in principle be used to teleport \cite{telep} one qubit. A simple protocol to teleport one qubit encoded in a superposition of coherent states $|u\rangle=|\alpha\rangle$ and $|v\rangle=|-\alpha\rangle$ was given in \cite{enk}: in order to perform the joint (Bell) measurement all the sender has to do is split the state to be teleported on a 50/50 beamsplitter with one half of the entangled coherent state, and count photons in the two output ports. If an odd number of photons is found in any one of the outputs, then perfect (unit fidelity) teleportation is possible\cite{note1}. The probability to find an odd number of photons turns out to be 50\%. 

Later a very similar protocol was found to work for entangled squeezed states as well \cite{cai}. That is, if $|u\rangle$ and $|v\rangle$ are squeezed vacuum states with opposite phases the sender still can apply the same joint measurement, and again an odd number of photons in either output port leads to perfect teleportation. Here, however, the probability to detect an odd number of photons is only 25\%.

Here we show that those similarities are no coincidence and are in fact a general feature of the states (\ref{ent1}), independent of the properties of coherent states or squeezed states, and regardless of whether $|u\rangle$ and $|v\rangle$ have opposite phases.

We also note that a recent experiment\cite{martini} (see also \cite{boas}) succeeded in teleporting a qubit encoded in superpositions of the vacuum $|0\rangle$ and a single-photon state $|1\rangle$ of a particular EM field mode, by making use of the state
\begin{equation}\label{01}
|\Phi\rangle=(|0\rangle|1\rangle-|1\rangle|0\rangle)/\sqrt{2}.
\end{equation}
Here the joint measurement again consists of splitting the state to be teleported on a 50/50 beamsplitter with half of the entangled state (\ref{01}); and again, the detection of a single photon (where 1 is an odd number) in any of the output ports corresponds to projecting onto one of the Bell states. 
The similarity with the above is easy to explain, as the protocol is in fact identical to that for entangled coherent states, in the limit that the coherent state amplitude $\alpha$ goes to zero. The success probability is, therefore, 50\%\cite{cals}. 
\section{Two simple facts}
Before stating two simple facts that turn out to be useful for understanding the above-mentioned teleportation protocols, we
need some preliminary results.

Consider a state of two EM field modes $A$ and $B$ of the following form
\begin{equation}\label{K}
|\Psi\rangle_{A,B}=\sum_{n,m} K_{nm} \frac{(B^{\dagger}+iA^{\dagger})^n}{\sqrt{n!}(\sqrt{2}^n)}
\frac{(B^{\dagger}-iA^{\dagger})^m}{\sqrt{m!}(\sqrt{2}^m)}
|0\rangle_A|0\rangle_B,
\end{equation}
with $A^{\dagger}$ and $B^{\dagger}$ the creation operators of modes $A$ and $B$.
If $K_{nm}$ is symmetric in its indices $(n,m)$ then it is easy to verify that 
only even powers of $A^{\dagger}$ occur with nonzero amplitude. In other words, the number of photons in mode $A$ is always even. On the other hand, if $K_{nm}$ is antisymmetric, then only odd powers of $A^{\dagger}$ survive, and mode $A$ always has an odd number of photons. In general, if one decomposes $K_{nm}$ in its symmetric and antisymmetric parts, $K_{nm}=K^s_{nm}
+K^a_{nm}$, then the probability to find an odd number of photons in mode $A$ is
\begin{equation}\label{Podd}
P^A_{{\rm odd}}=\frac{\sum_{nm}|K^a_{nm}|^2}{\sum_{nm}|K_{nm}|^2}.
\end{equation}
(Note $\sum_{nm}|K_{nm}|^2=\sum_{nm}|K_{nm}^a|^2+\sum_{nm}|K_{nm}^s|^2$.)
We also need
the unitary phase shift operator: it is given by
\begin{equation}
U_\phi=\exp(i\phi a^\dagger a)
\end{equation}
for an EM field mode
described by creation and annihilation operators $a^\dagger$ and $a$.
\newtheorem{fact}{Fact}
\begin{fact}
If one splits a pure state $|\psi\rangle$ and its phase-shifted version
 \begin{equation}
|\tilde{\psi}\rangle\equiv U_{\pi/2}|\psi\rangle
\end{equation}
on a 50/50 beamsplitter, one of the two output ports never contains an odd number of photons. 
\end{fact}
Proof: Expand $|\psi\rangle$ and $|\tilde{\psi}\rangle$ in photon-number states $|n\rangle$ as
\begin{eqnarray}
|\psi\rangle&=&\sum_n p_n|n\rangle,\nonumber\\
|\tilde{\psi}\rangle&=&\sum_n i^n p_n|n\rangle.
\end{eqnarray}
Then we use that for a 50/50 beamsplitter 
the creation operators of the two output ports are given by
\begin{eqnarray}
A^{\dagger}&=&\frac{a^{\dagger}+ib^{\dagger}}{\sqrt{2}}\nonumber\\
B^{\dagger}&=&\frac{ia^{\dagger}+b^{\dagger}}{\sqrt{2}}
\end{eqnarray}
in terms of the creation operators of the two input modes.
Thus, starting with a state $|\tilde{\psi}\rangle_a|\psi\rangle_b$ for input modes $a$ and $b$ 
a 50/50 beamsplitter transforms that state into  
\begin{eqnarray}
&&|\tilde{\psi}\rangle_a|\psi\rangle_b\longrightarrow\nonumber\\
&&\sum_n p_n \frac{(B^{\dagger}+iA^{\dagger})^n}{\sqrt{2}^n\sqrt{n!}}
\sum_m p_m  \frac{(B^{\dagger}-iA^{\dagger})^m}{\sqrt{2}^m\sqrt{m!}}|0\rangle_A|0\rangle_B.
\end{eqnarray}
We now have a state of the form (\ref{K}) with symmetric coefficients $K_{nm}=p_np_m$, so that mode $A$ always contains an even number of photons.
(Obviously, if the inputs of modes $a$ and $b$ are interchanged, it is mode $B$ that always contains an even number of photons).
\begin{fact}
 Take any pure state $|\psi\rangle$ and any pure state $|\phi\rangle$ orthogonal to $|\psi\rangle$. If one splits $|\psi\rangle$ and
$|\tilde{\phi}\rangle=U_{\pi/2}|\phi\rangle$
on a 50/50 beamsplitter, one of the two output ports contains an odd number of photons with exactly 50\% probability. 
\end{fact}
Proof: Expand $|\psi\rangle$ and $|\tilde{\phi}\rangle$ in photon-number states
\begin{eqnarray}
|\psi\rangle&=&\sum_n p_n|n\rangle,\nonumber\\
|\tilde{\phi}\rangle&=&\sum_n i^n q_n|n\rangle,
\end{eqnarray}
with
\begin{equation}\label{ortho}
\sum_n p_n q_n^*=0.
\end{equation}
The state that results after splitting these two states on a 50/50 beamsplitter
is again of the form (\ref{K}) with
$K_{nm}=p_nq_m$. 
Because of the relation
\begin{equation}
\sum_{nm}|p_nq_m+p_mq_n|^2=\sum_{nm}|p_nq_m-p_mq_n|^2,
\end{equation}
which follows immediately from (\ref{ortho}),
the symmetric and antisymmetric parts of $K_{nm}$ satisfy
\begin{equation}
\sum_{nm}|K^a_{nm}|^2=\sum_{nm}|K^s_{nm}|^2.
\end{equation}
According to (\ref{Podd}) we have  $P^A_{{\rm odd}}=1/2$.
\section{Teleportation}\label{T}
We apply the previous results to construct a simple teleportation protocol involving the states (\ref{ent1}).
Let
\begin{eqnarray}
|u\rangle&=&\sum_n a_n |n\rangle,\nonumber\\
|v\rangle&=&\sum_n b_n |n\rangle,
\end{eqnarray}
with
\begin{mathletters}
\begin{eqnarray}
\sum_n |a_n|^2&=&\sum_n |b_n|^2=1,\\
\sum_n a_n b^*_n&=&\sum_n a^*_nb_n,\label{cond}
\end{eqnarray}
\end{mathletters}
be two arbitrary states of the EM field with real inner product.
Then the two superpositions $|\pm\rangle\sim |u\rangle \pm |v\rangle$,
\begin{eqnarray}
|+\rangle&=&\sum_n c_n^+ |n\rangle,\nonumber\\
|-\rangle&=&\sum_n c_n^- |n\rangle,
\end{eqnarray}
with
\begin{eqnarray}\label{c+-}
c_n^+=\frac{(a_n+b_n)}{\sqrt{\sum_n |a_n+b_n|^2}}, \nonumber\\
c_n^-=\frac{(a_n-b_n)}{\sqrt{\sum_n |a_n-b_n|^2}},
\end{eqnarray}
are orthogonal. Hence, the states
\begin{eqnarray}\label{ent+}
|\Psi_-\rangle&=&(|+\rangle|-\rangle+|-\rangle|+\rangle)/\sqrt{2},\nonumber\\
|\Phi_-\rangle&=& (|-\rangle|+\rangle-|+\rangle|-\rangle)/\sqrt{2},
\end{eqnarray}
manifestly have one ebit of entanglement and are in fact equal to the states defined in (\ref{ent1}), but this time properly normalized.
\subsection{The basic protocol}
Take an arbitrary qubit in mode $a$, encoded in the phase-shifted versions of $|\pm\rangle$ as  
\[ \epsilon_+ |\tilde{+}\rangle_a +\epsilon_- |\tilde{-}\rangle_a, \]
and combine it with one half of the state $|\Phi\rangle$ from Eq.~(\ref{ent+}), that we take to be a state of modes $b$ and $c$, on a 50/50 beamsplitter. 
This leads to the transformation
\begin{eqnarray}\label{eps}
&&[\epsilon_+ |\tilde{+}\rangle_a +\epsilon_- |\tilde{-}\rangle_a] 
[|-\rangle_b|+\rangle_c-|+\rangle_b|-\rangle_c]/\sqrt{2}\longrightarrow\nonumber\\
&&\sum_{nm} \frac{(B^{\dagger}+iA^{\dagger})^n}{\sqrt{2}^n\sqrt{n!}}
\frac{(B^{\dagger}-iA^{\dagger})^m}{\sqrt{2}^m\sqrt{m!}}|0\rangle_A|0\rangle_B
\nonumber\\
&&\{\epsilon_-c_n^-c_m^-|+\rangle_c
-\epsilon_+c_n^+c_m^+|-\rangle_c  \nonumber\\
&&+\epsilon_+c_n^+c_m^-|+\rangle_c
-\epsilon_-c_n^-c_m^+|-\rangle_c  \}/\sqrt{2}.
\end{eqnarray}
By Fact 1 of the preceding Section the combinations $|+\rangle_a |\tilde{+}\rangle_b$ and
$|-\rangle_a |\tilde{-}\rangle_b$ never lead to an odd number of photons in mode $A$. Thus, upon detection of an odd number $N_A$ of photons in mode $A$ one projects onto the remaining two terms, corresponding to the two terms in the last line of Eq.~(\ref{eps}). Now, as noted before, the terms with odd powers of $A^{\dagger}$  correspond to terms antisymmetric in $(n,m)$. Hence, replacing $c_n^-c_m^+$ with $-c_m^-c_n^+$ in the last line directly shows that the state of mode $c$ is reduced to
\[
\epsilon_+ |+\rangle_c +\epsilon_-|-\rangle_c.
\]
If one wishes one may perform the unitary transformation that connects $|\pm\rangle$ to $|\tilde{\pm}\rangle$, which is just a phase shift over $\pi/2$, to complete teleportation from mode $a$ to mode $c$. 

The probability to find an odd number of photons in mode $A$, and hence to succeed in perfect teleportation, is independent of the state to be teleported. Moreover, it is independent of the states $|u\rangle$ and $|v\rangle$, as the probability is
\begin{equation}
P_{{\rm success}}=\frac{1}{4}.
\end{equation}
One factor 1/2 arises from having to filter out the $|+\rangle_a |\tilde{+}\rangle_b$ and
$|-\rangle_a |\tilde{-}\rangle_b$ combinations, the other factor 1/2 arises from Fact 2. 
This success probability of 1/4 is indeed the probability found for the teleportation protocol discussed in \cite{cai}, using squeezed vacuum states for $|u\rangle$ and $|v\rangle$.
\subsection{Increasing the success probability}
For arbitrary states $|u\rangle$ and $|v\rangle$ other measurement outcomes (in particular, finding an odd number of photons in mode $B$) do not in general lead to perfect teleportation.
However, by choosing the states wisely one can increase the probability of successful teleportation to 50\%.

Suppose we choose $|u\rangle$ and $|v\rangle$ such that
\begin{equation}\label{wise1}
|v\rangle=U_{\pi}|u\rangle.
\end{equation}
The condition that the inner product between $|u\rangle$ and $|v\rangle$ be real, Eq.~(\ref{cond}), is automatically satisfied then. 
Moreover, the coefficients $c_n^+$ of Eq. (\ref{c+-}) are nonzero only for $n$ even, while $c_n^-$ can be nonzero only for odd values of $n$.
This implies that the combinations $|+\rangle_a |\tilde{+}\rangle_b$ and
$|-\rangle_a |\tilde{-}\rangle_b$ always lead to an even number of photons not only in mode $A$, as before, but in mode $B$ as well. On the other hand, the other two combinations, $|+\rangle_a |\tilde{-}\rangle_b$ and
$|-\rangle_a |\tilde{+}\rangle_b$ always lead to an odd number of photons in modes $A$ and $B$ together. Both possibilities occur with the same probability of 1/4, and in both cases perfect teleportation is possible. If the odd number of photons is detected in mode $A$ then we have the same situation as before, if the odd number is detected in mode $B$, then inspection of Eq.~(\ref{eps}) reveals that mode $c$ ends up as
\[
\epsilon_+ |+\rangle_c -\epsilon_-|-\rangle_c.
\]
Again only a trivial unitary transformation is needed to complete teleportation from mode $a$ to mode $c$.

Hence, for states of the form (\ref{wise1}) one just needs to detect an odd number of photons in either mode (in which case the other mode has an even number of photons), in order to achieve perfect teleportation. The success probability is
\begin{equation}
P_{{\rm success}}=\frac{1}{2}
\end{equation}
independent of the state to be teleported and independent of the choice of $|u\rangle$.
Both the teleportation protocol with entangled coherent states as described in \cite{enk} and the protocol used in \cite{martini} fit into this category. 
For in the former protocol $|u\rangle=|\alpha\rangle$ and 
$|v\rangle=U_\pi|u\rangle=|-\alpha\rangle$ for coherent states $|\alpha\rangle$, while in the latter one may choose $|u\rangle=(|0\rangle+|1\rangle)/\sqrt{2}$ and $|v\rangle=U_\pi|u\rangle=(|0\rangle-|1\rangle)/\sqrt{2}$.
\subsection{Quantum scissors}
Apart from the three teleportation protocols mentioned in the Introduction, other protocols can now be constructed along the same lines by choosing particular states $|u\rangle$ and $|v\rangle$. 
For instance, by splitting two single-photon states on a 50/50 beamsplitter one generates a state of the form
\begin{equation}\label{02}
|\Phi\rangle=(|0\rangle|2\rangle-|2\rangle|0\rangle)/\sqrt{2},
\end{equation}
which clearly is of the correct form (\ref{ent1}). Instead of analyzing teleportation, we consider the slightly more general quantum scissor\cite{scissor} protocol, in which states are teleported in truncated form. 

Start with an arbitrary state of mode $a$, $\sum_n\alpha_n i^n|n\rangle_a$, and the state (\ref{02}) of modes $b,c$, and as usual combine them on a 50/50 beamsplitter:
\begin{eqnarray}
&&\sum_n \alpha_n i^n|n\rangle_a  (|0\rangle_b|2\rangle_c-|2\rangle_b|0\rangle_c)/\sqrt{2}\longrightarrow\nonumber\\
&&\sum_n \alpha_n
\frac{(B^{\dagger}+iA^{\dagger})^n}{\sqrt{2}^{n+1}\sqrt{n!}}|0\rangle_A|0\rangle_B
|2\rangle_c\nonumber\\
&&-\sum_n \alpha_n
\frac{(B^{\dagger}+iA^{\dagger})^n(B^{\dagger}-iA^{\dagger})^2}{\sqrt{2}^{n+4}\sqrt{n!}}|0\rangle_A|0\rangle_B
|0\rangle_c.
\end{eqnarray}
Now suppose one finds $N_A$ photons in mode $A$ and $N_B$ in mode $B$.
Mode $c$ is then collapsed into the (unnormalized: the norm gives the probability to have this many photons in the modes $A,B$) state 
\begin{eqnarray}
&&\alpha_{N_t}
\left(
\begin{array}{c}
N_t\\N_B
\end{array}\right)\frac{\sqrt{N_B!N_A!}}{\sqrt{N_t!}}\frac{i^{N_A}}{\sqrt{2}^{N_t+1}}|2\rangle_c\nonumber\\
&-&
\alpha_{N_t-2}
\left(
\begin{array}{c}
N_t-2\\N_B-2
\end{array}\right)\frac{\sqrt{N_B!N_A!}}{\sqrt{(N_t-2)!}}\frac{i^{N_A}}{\sqrt{2}^{N_t+2}}|0\rangle_c\nonumber\\
&+&
\alpha_{N_t-2}
\left(
\begin{array}{c}
N_t-2\\N_B
\end{array}\right)\frac{\sqrt{N_B!N_A!}}{\sqrt{(N_t-2)!}}\frac{i^{N_A-2}}{\sqrt{2}^{N_t+2}}|0\rangle_c\nonumber\\
&+&2i
\alpha_{N_t-2}
\left(
\begin{array}{c}
N_t-2\\N_B-1
\end{array}\right)\frac{\sqrt{N_B!N_A!}}{\sqrt{(N_t-2)!}}\frac{i^{N_A-1}}{\sqrt{2}^{N_t+2}}|0\rangle_c,
\end{eqnarray}
where $N_t=N_A+N_B$ the total number of photons in modes $A,B$.
Two cases are of interest here: if $N_A=N_B=1$, then the state of mode $c$ is collapsed into
\[\alpha_0|0\rangle_c+\alpha_2|2\rangle_c\]
with probability $P=[|\alpha_0|^2+|\alpha_2|^2]/4$. Similarly, 
if $N_A=2,N_B=0$ or $N_A=0,N_B=2$, one gets
\[\alpha_0|0\rangle_c-\alpha_2|2\rangle_c\]
with the same probability $P=[|\alpha_0|^2+|\alpha_2|^2]/4$. Thus, all coefficients of the original state of mode $a$ are cut away, except for $\alpha_0$ and $\alpha_2$, and the truncated state ends up in mode $c$.  

This scheme can be generalized in an obvious way. By following the same steps
but starting with a different entangled state,
\begin{equation}\label{NM}
|\Psi\rangle=(|N\rangle|M\rangle-|M\rangle|N\rangle)/\sqrt{2},
\end{equation}
with $M\neq N$, an arbitrary state $\sum_n\alpha_ni^n|n\rangle_a$ is turned into 
\[\alpha_N|N\rangle_c+(-1)^{N_A}\alpha_M|M\rangle_c,\]
provided one detects exactly $N+M$ photons in total in modes $A,B$. 
This occurs with probability $P=[|\alpha_N|^2+|\alpha_M|^2]/2$. 
The unitary operation that has to be performed to complete the protocol depends on whether the number of photons detected in mode $A$ is odd or even.
\section{Discussion}
Any state of the form
$|u\rangle|v\rangle-|v\rangle|u\rangle$ possesses one ebit of entanglement\cite{hirota}. As such, it is no surprise it can be used to teleport one qubit encoded in an arbitrary superposition of $|u\rangle$ and $|v\rangle$. What is surprising, is that there is a simple teleportation procedure that does not depend on what the states $|u\rangle$ and $|v\rangle$ are, as long as they are pure states of the electromagnetic field:  Split the state to be teleported on a 50/50 beamsplitter with half of the entangled state, and count photons  in one specific output port. If the number is odd, one has teleported the state from one mode to another. In general, the probability to find an odd number of photons in that particular output is 25\%. However, by choosing the state $|v\rangle$ equal to $|u\rangle$ up to a phase shift over an angle $\pi$, finding an odd number in the other output mode leads to teleportation as well, increasing the probability of successful teleportation to 50\%.

For a special subclass of entangled states where $|u\rangle=|N\rangle$ and $|v\rangle=|M\rangle$ are different number states we also showed how to perform a generalized quantum scissors\cite{scissor} protocol; an arbitrary state is teleported in truncated form, where only the two terms containing exactly $N$ and $M$ photons, respectively, survive. 

For these protocols one needs measurements that distinguish odd from even numbers of photons. Interestingly, but perhaps not surprisingly, such measurements also can be used to detect nonlocal entanglement in the type of states considered here\cite{Bell,Bellc} by testing Bell inequalities.
And similarly, such measurements can be used to purify mixed entangled coherent states\cite{puri}.
Experimentally, this is not a straightforward task, as the best detectors
have an efficiency of about 85\% in the optical frequency domain\cite{detector}.
This implies that the distinction between even and odd numbers of photons is only reliable in practice for low photon numbers.


\begin{thebibliography}{99}
\bibitem{note1}Of course, perfect teleportation is achieved only in the absence of noise. In particular, photon detectors are assumed to have unit efficiency here. See also the Discussion Section.
\end{thebibliography}
\end{document}